\documentclass[11pt]{article}
\usepackage[latin1]{inputenc}
\usepackage[english]{babel}
\usepackage[namelimits]{amsmath}
\usepackage{amssymb}
\usepackage{amsmath}
\usepackage{amsthm}

\begin{document}
\title{Bekenstein-Hawking entropy in expanding universes from black hole theorems}
\author{
Stefano Viaggiu,\\
Dipartimento di Matematica,
Universit\`a di Roma ``Tor Vergata'',\\
Via della Ricerca Scientifica, 1, I-00133 Roma, Italy.\\
E-mail: {\tt viaggiu@axp.mat.uniroma2.it}}
\date{\today}\maketitle
\begin{abstract}
We show that the use of suitable theorems for black hole formation in Friedmann expanding universes
leads to a modification of the Bekenstein-Hawking entropy. By adopting an argument similar to the original Bekenstein one,
we write down the expression for the  Bekenstein-Hawking entropy suitable for non-static isotropic expanding universes together with the equation of state of a black hole. This equation can be put in a form similar to the one of an ideal gas but with a factor depending on the Hubble radius. Moreover, we
give some argument on a possible relation between our entropy expression
and the Cardy-Verlinde one. 
Finally, we explore the possibility that primordial inflation is due to black hole evaporation in our context.
\end{abstract}

\section{Introduction}
The  Hawking's discovery that black holes \cite{Haw1} have thermal radiations gave a fundamental tool to investigate 
black hole thermodynamics in general relativity. In fact, in a asymptotically flat spacetime, to a static spherically symmetric black hole of 
proper area $A$ can be associated an entropy $S_{BH}$ given by $S_{BH}=\frac{k_B A}{4 L_P^2}$, where
$k_B$ is the Boltzmann constant and $L_P$ the Planck length. 
The situation is much more involved in a cosmological non static background.
As an example, only recently has been proved that 
\cite{kal} the McVittie solution \cite{mac}  
contains a black hole in an expanding universe. Moreover, it has been shown \cite{H1,H2}
that in an expanding universe, an important ingredient is provided by the apparent horizon for a black hole, 
in particular for the definition of its entropy. Other examples of study of entropy of black holes embedded in an expanding universe can be found, for example in
\cite{E1,E2,E3}. However, to the best of my knowledge, 
in all these papers one assumes for the entropy $S_{BH}$ the same expression of the static  background, i.e.  
$S_{BH}=\frac{k_B A}{4 L_P^2}$,
where $A$ denotes the proper area of the apparent horizon.
In this paper,  by considering suitable theorems for the formation of trapped surfaces in expanding universes \cite{13,17},
we explore the possibility  that the expression for the black hole entropy in expanding universes should contain an extra term depending on the Hubble flow. 
In section 2 we present the black hole theorems suitable for an expanding universe. 
In section 3 we study the consequences of our new 
expression for $S_{BH}$. 
Section 4 is devoted to a discussion of the
entropy bounds, while
section 5 we preliminarly study the inflationary paradigm within our proposal.   
Finally, section 6 is devoted to some conclusions and final remarks.

\section{Black Holes Theorems in Flat Expanding Universes}

An important issue in general relativity concerns the conditions allowing a mass-energy concentration in a certain volume
to collapse in a black hole. On large scale, our universe 
is well approximated by a spatially flat Friedmann metric, so in this paper we 
mainly consider this case.

In the 
case of spherical collapse in asymptotically flat background, if the condition  
$2GM/c^2>R$ ($R$ is the radius of the sphere and $M$ the ADM total mass-energy) is fulfilled, a black hole arises. However $M$ is the total energy (together with the negative binding energy) not the one effectively concentred within $R$ and $R$ itself is not an appropriate measure of the proper volume inside $R$. A fundamental ingredient characterizing a black hole is its event horizon. Unfortunately,
the identification of the event horizon is a global property of the spacetime and its identification is a complicated task. A 
more manageable (local)  ingredient is provided by trapped surfaces. In fact in \cite{isr} it has been shown that the formation of trapped surfaces (TS)
caused by spherically symmteric mass-energy concentration satisfying the weak energy condition, unavoidably leads to a 
black hole. In a series of papers \cite{13,17} one can find necessary conditions for the non formation of TS and sufficient conditions for the formation of TS expressed only in terms of proper quantities in the spherical case. In particular in 
\cite{14,17} such conditions are done in Friedmann cosmologies.\\
In this paper, we consider the theorem shown in \cite{17} for open Friedmann flat cosmologies.

Consider, in a flat Friedmann cosmology with a background energy-density ${\overline{\rho}}_m$,
a spherical two surface $S$ of proper radius $L$ and proper area $A$ and
a perturbation of proper mass $\delta M>0$ within $S$.
Initial conditions are done in a three dimensional manifold $\Sigma$ that can be seen as a spacelike slice of the four dimensional manifold representing the solution of the Einstein's equations. The two-surface $S$ is embedded in $\Sigma$.
Suppose that the current matter perturbation $\delta J_{\mu}$
is vanishing on the boundary of $S$, i.e. that the matter is at rest on $S$ and that 
the trace of the extrinsic curvature $k_{\alpha\beta}$ is constant on comoving foliation ($t=const.$), i.e. $k^{\alpha}_{\alpha}=const.$
As a consequence, if
\begin{equation}
\delta M\frac{G}{c^2}<\frac{L}{2}+A\sqrt{\frac{G\;{\overline{\rho}}_m}{6\pi c^2}},
\label{3}
\end{equation}
then $S$ is not trapped. 
Note that $L$ and $A$
are respectively the proper length and the proper area with respect to 
the perturbed configuration, i.e. backreaction is taken into account. Moreover,
the upper bound for the mass-excess depends on the energy-density of the background on which the
perturbation acts. The authors of \cite{17} do not quote necessary and sufficient conditions for the formation of TS. Fortunately, in 
\cite{17} we can found a necessary condition for the formation of TS.
Under the same conditions of (\ref{3}),  if
\begin{equation}
\delta M\frac{G}{c^2}>L+A\sqrt{\frac{G\;{\overline{\rho}}_m}{6\pi c^2}},
\label{3bis}
\end{equation}
then $S$ is trapped. Therefore, by introducing a real constant $\gamma\in[1,2]$, we can argue from 
(\ref{3}) and (\ref{3bis}) that the necessary and sufficient condition for the non formation of TS is
\begin{equation}
\delta M\frac{G}{c^2}<\frac{\gamma L}{2}+A\sqrt{\frac{G\;{\overline{\rho}}_m}{6\pi c^2}}.
\label{3tris}
\end{equation}
In any case, with
respect to the static case, a further term  appears proportional to $\sim A\sqrt{{\overline{\rho}}_m}$. This term can also be written
as a function 
of the cosmological constant $H$. In fact, if all the energy-densities present in the universe $\sum_i{\rho}_i$ 
satisfy the Friedmann equation 
$H^2=8/3\pi G\sum_i{\rho}_i$ (i.e. we have a spherical black hole embedded in a Friedmann expanding universe, see
\cite{wit, mac}), then inequality (\ref{3tris}) becomes:
\begin{equation}
\delta M\frac{G}{c^2}<\frac{\gamma L}{2}+\frac{AH}{4\pi c}.
\label{3z}
\end{equation}
The condition (\ref{3z}) indicates, according to physical intuition, that an expanding universe makes more difficult the formation of trapped surfaces  and as a consequence the formation of black holes \cite{16}. The interesting feature of the inequality (\ref{3z}) it 
is that is expressed only in terms of proper quantities that are in principle measurable.\\
The theorem quoted above is the starting point to investigate its consequences on the maximal entropy allowed in a certain spacetime region. It contains only well defined proper local quantities. 

\section{Generalized Bekenstein-Hawking Entropy}
An interesting first consequence of the reasonings above is the modification of the Bekenstein-Hawking entropy. 
In fact, if the maximum entropy contained in a spherical region of radius $R$ and energy $E$
is provided by the entropy of a black hole of the same 
radius, then a correction term due to $H$ must be included. Otherwise, 
a certain region of radius $R$ should contain 'more entropy' than the one used for a black hole, i.e. $S_{BH}=\frac{k_B A}{4 L_P^2}$,
but without forming, thanks to the inequality (\ref{3z}), a black hole: a contradiction.\\
In this section we explore this interesting aspect of our proposal.

\subsection{Bekenstein bound and entropy bound}

A first interesting consequence of our result is that 
expression (\ref{3}) suggests a correction term caused by the degrees of freedom that are due to the 
non-static nature of Friedmann spacetimes. To start with,
consider the old original
Bekenstein argument \cite{Bek} in an asymptotically flat spacetime in a static context. 
This argument states that there exists an universal bound for the
entropy $S$ of a spherical object of radius $R$ and energy $E$ given by 
$S\leq S_{max}=\frac{2\pi k_B RE}{\hbar c}$.\\ 
The derivation of this bound originated a dispute (see for example \cite{d1,d2,d3,d4}). This bound can be 
violated in the case of strong gravitational fields as a system collapsed inside a black hole 
(although the metric inside the event horizon becomes time dependent and calculations are not so obvious)
and in a cosmological context (see \cite{d5} for a proposal to general spacetimes in order 
to save the holographic principle). This discussion, as we see below, although important it is not essential for our purposes to give a 
more general expression for the black hole entropy in expanding universe.
In any case, an improvement of the Bekenstein original bound has been done by Susskind (see \cite{bb1,bb2}). The argument is the following.
Since $E\sim R$ and more precisely $E\leq E_{max}=c^4 R/(2G)$, 
the  Bekenstein bound results proportional to the proper area of the object, i.e. 
$S_{max}= \frac{k_B A}{4 L_P^2}$. Hawking calculations \cite{Haw1} confirmed
this factor to be $1/4$ leading to the  well known expression for the black hole entropy by showing that this bound is satured by 
the black holes. This bound (spherical entropy bound) is weaker than the Bekenstein one. It is a consequence of the requirement
that a system be gravitationally  stable, otherwise, a gravitational collapse with a final
black hole arises. Also this bound fails in practical 
situations as the case of a collapsing object. A further improvement
is provided by the spacelike bound (see \cite{bb2} and references therein).
This bound implies that the entropy of the matter enclosed in a 
compact spacelike surface of volume $V$ and area $A$ never exceeds the entropy of a black hole  
of area $A$. Also in this form (see the discussion at section 4), counterexamples can be found.

The situation is much more involved in a cosmological dynamical context. There, because of the non static nature of the spacetime,
the bound becomes dynamical \cite{bb3,bb4}. In particular, in \cite{bb3} the holographic principle is discussed in 
the context of
a closed radiation dominated Friedmann universe where the radiation is represented by a conformal field theory with a suitable central charge.
In this frame emerged a new bound satured when the Cardy formula is used. This bound depends on the Hubble flow
$H$ (see the discussions at  subsection 3.3). The most interesting bound according to the holographic principle working also in  cosmological spacetimes is the one quoted by Bousso \cite{d5,bb2}. There, the bounding area is not the one enclosed in a 
region $B$ but rather the one expressed in terms of light-sheet of $B$ itself. This bound seems to work but its physical origin
is still obscure. 

Summarizing, the reasonings above show that, 
in order to save the holographic principle, the area appearing in the right hand  side of the spacelike
entropy bound must be greater than the area enclosing the surface $A$. This is an important lesson of the Bousso idea. 

In this paper we are mainly interested in a generalization of the Bekenstein-Hawking entropy in a cosmological context. In this regard, it is sufficient to consider the spacelike entropy bound. In fact, in this way we can obtain as an upper  bound the correct expression for the
black hole entropy in asymptotically flat spacetime. Then, with the use of theorem (\ref{3z}) we can infer the new form of the black hole entropy suitable in expanding universes.

\subsection{Bekenstein-Hawking entropy in expanding universes}

Following the entropy bound argument, we propose a generilized expression for the Bekenstein-Hawking formula suitable 
in a flat expanding universe with embedded a black hole.
To our purposes, 
it is more appropriate to work with proper distances. There, $E$ will denote the proper energy within $S$ and $R$
the proper length $L$ of the sphere. 
Thanks to (\ref{3z}), the maximal allowed energy $E_{max}$  in a spherical region of proper radius $L$ above which a black hole arises is  
\begin{equation}
E_{max}=\frac{c^4}{G}\left[\frac{\gamma L}{2}+A\sqrt{\frac{G\;{{\rho}}_m}{6\pi c^2}}\right],
\label{bo1}
\end{equation}
where ${\rho}_m$ denotes the matter-energy content of the universe. 
It is important to note that the proper quantities $L$ and $A$ in (\ref{bo1})
refer to the one of a black hole.
The identification
of the surface $A$ with the analogue of the event horizon in the Schwarzschild case
in expanding universes is not (see for example \cite{wit}) a simple task.  
Since we are considering black holes in expanding universes, 
the enclosing surface $A$, in light, for example, of the works \cite{H1,H2} can be well identified with the  
apparent horizon of the black hole. 
It is not 
obvious how to define the proper volume or the proper length of a black hole, since the radial coordinate is time-like inside $V$.
As shown in \cite{vol}, one can define, for non rotating black holes,  
an effective volume (named geometric volume)
as minus the volume excluded from a spatial slice by the black hole horizon, i.e. $V_h=4/3\pi R_h^3$ 
where $R_h=R(t,r_h)$ is the scale factor in a spherically symmetric expanding universe (i.e. the the angular part of the metric, 
at $t=const$ and $R=const$, can be always written in the form $R^2(d{\theta}^2+{\sin}^2\theta d{\phi}^2)$)
and $r_h$
measures the comoving radius of the apparent horizon of the embedded black hole. 
According to the definition of $V_h$, we can define the effective proper length of the black hole as $L_h=R_h$. In this context
$H=R_{,t}/R$. In fact, in the original derivation of theorem (\ref{3}), the authors 
of \cite{14,15,17} work in isotropic coordinates
by factorizing $R$ as $R={\phi}^2(r)a(t)$. In this way we have that $H=R_{,t}/R=a_{,t}/a$, as happens for example for the McVittie solution
\cite{mac}.\\
The factor $\gamma$ can be fixed from the fact that in the static limit, for the Bekenstein bound we obtain:
\begin{equation}
S_{max}=S_{BH}=\frac{k_B\gamma A_h}{4 L_P^2},
\label{static}
\end{equation}
and thus $\gamma=1$.
As a consequence of the above reasonings, 
with
$L_h=\sqrt{\frac{A_h}{4\pi}}$,
for the entropy $S_{BH}$ we obtain 
\begin{equation}
S_{BH}=\frac{k_B A_h}{4 L_P^2}+\frac{k_BA_h^{\frac{3}{2}}}{c L_P^2}\sqrt{\frac{G{\rho}_m}{6}}.
\label{entr}
\end{equation}
The correction term takes into account
the degrees of freedom of a non-static (flat) Friedmann universe. 
Expression (\ref{entr}) can also be written in terms of $H$ instead of ${\rho}_m$:
\begin{equation}
S_{BH}=\frac{k_B A_h}{4 L_P^2}\left(1+\frac{H}{c}\sqrt{\frac{A_h}{\pi}}\right).
\label{entr2}
\end{equation}
The formula (\ref{entr2}) says that our bound is greater than the ones given by the first term only:
an higher admissible upper bound for the density implies an higher upper bound for the entropy.
The usual term of the black hole entropy, i.e. the first one in the right hand side of (\ref{entr2}), derives on the term $L/2$ in
(\ref{3}). Consider now a black hole in an expanding universe with apparent horizon area $A_h$. If the usual expression
for the entropy $S_u$ where $S_u=\frac{k_B A_h}{4 L_P^2}$, thanks to (\ref{3z}) we could have an object with entropy $S$
such that $S_u<S<S_{BH}$, i.e. an object with an entropy greater than the one of a black hole of the same size but without forming a black hole, because in expanding universe more energy is required in a spherical surface A to form a TS. This is a logic contradiction 
that is solved by adding to the black hole entropy the volume term $\sim A_h^{3/2}H$. 
The bound $S<S_{BH}$ is an entropy bound. In subsection 3.3  a new interpretation of this bound, in light of the theorems used
in section 2, is done
solving some problems of the original formulation.\\
Generally, the correction term is negligible when $A_h<<c^2/H^2$: when the proper dimensions of an object become comparable with
the Hubble radius, this correction cannot be neglected. As a title of example,
suppose to have a spherical region of comoving radius $r_c$ with 
a power low cosmology $a(t)=a_0 t^{\alpha},\;\alpha>0$ (in this case $R=L=a(t) r_c$).
The correction term in (\ref{entr2}) becomes dominant when
$r_c>>c/(a_{,t})=ct^{1-\alpha}/(a_0\alpha)$. As a consequence,  if the strong energy condition is fulfilled ($\alpha<1$), 
the correction term can dominate near the big bang singularity. Conversely, for inflationary cosmologies $\alpha>1$ with a
power law behaviour, the added term
becomes dominant also for relatively small comoving regions, provided that a sufficient cosmic time is 
considered: this phenomenon also appears for a universe filled with a cosmological constant where asymptotically we have
$H\rightarrow c\sqrt{\Lambda/3}$. Also during the primordial quasi de Sitter inflationary era, this correction term can be relevant.  

It is interesting to study the fate of the full expression
(\ref{entr2}) near the big bang. The first term is always vanishing at $t=0$. Concerning the added term, 
if $H\sim 1/t$, then for a power law cosmologies with $a\sim t^{\alpha}$, for $\alpha\in (0, 1/3)$, 
$S_{BH}\rightarrow\infty$. Conversely, for a spacetime  with $\alpha>1/3$,  
$S_{BH}\rightarrow 0$. Interestingly enough, for stiff matter, i.e. $\alpha=1/3$, then entropy reaches at the big bang a finite non-vanishing limit. The fact that only for superluminar acausal fluids the entropy diverges at the big bang 
strongly supports the physical reasonability of our proposal.

As a further remark, we analyze the expressions (\ref{entr}) and (\ref{entr2})  from the point of view of the first law of thermodynamics. 
To this purpose, note that
in (\ref{entr}) the term $A_h^{3/2}$ can be written also in terms of  $V$. 
Since we are in a spherically symmetric context,  in (\ref{entr})
we have firstly explicitally written this term as a function of the proper area.
In practice, for a sphere in a spatially flat metric, the proper quantities $A_h$ and $V_h$ are not independent.
However,  in ligth to the first law of thermodynamics, it seems more natural and useful to express the added term as a function of the 
proper volume of the apparent horizon of the black hole.\\
To this purpose, we can follow a practical rule: if we have a behaviour 
$L^a$  with $a$ a positive integer, then we must have   $L^a\sim A^bV^q\sim L^{2b+3q}$, with $b,q$ positive integers.
In the case under consideration, the first term in (\ref{entr}) is  trivially $A$ (i.e. $b=1, q=0$), while the second is 
$V$ (i.e $b=0,q=1$). In practice surfaces and volumes do appear with integer exponents.
We obtain, after dropping the subscript $h$:
\begin{equation} 
S_{BH}=\frac{k_B A}{4 L_P^2}+\frac{3k_B}{2c L_P^2}V H.
\label{termo}
\end{equation}
In the following we study some interesting consequences of this formula.

\subsection{Dynamical degrees of freedom and the Cardy-Verlinde formula}
By a first inspection of the expression (\ref{termo}), we note that the added term is proportional to $VH$. This reminds 
the expression for the entropy arising in the context of a conformal field theory representing a radiation with a given central charge $C$
\cite{bb3}, by the help of the Cardy formula. In this context, the entropy is given, in $3+1$ dimensions, by 
$S_H=k_B HV/(2cL_P^2)$. In the case considered in \cite{bb3} for a closed Friedmann cosmology, 
the normalized dynamical Bekenstein bound is
$S\leq S_B=2\pi k_B ER/(3c\hbar)$ . 
It is a simple matter to verify that with the normilized factor $1/3$, the 
(dynamical) term proportional to 
$H$ in (\ref{termo}) is exactly identical to the one found by Verlinde in \cite{bb3}. This is an interesting fact. Moreover, 
the first term in our entropy (\ref{termo}) can solve the issue present for the Bekenstein bound when 
one  considers a closed universe near its maximum radius, without considering the covariant Bousso bound. 
In this case, $H\simeq 0$ and $S_H\rightarrow 0$, but the entropy for a universe
near its maximum it is non-vanishing  and so the term $\sim A$ in (\ref{termo}) allows the bound to be valid. 
As pointed in \cite{bb3}, for a strongly self-gravitating universe ($R_u H>c$, being $R_u$ the radius of the universe), 
the usual Bekenstein bound is still valid for regions larger than
the Hubble radius.
Conversely, for a (closed) universe with radius $R_u$ smaller than the Hubble one, 
the added term only is not enough to make the bound
valid. In fact, our formula (\ref{termo}) implies that for a universe with a radius smaller than the Hubble one, the firs term 
in (\ref{termo}) cannot be neglected. 
As an intriguing suggestion, thanks to the reasonings above,
the term $\sim VH$ in (\ref{termo}) could be seen as due to the manifestation of
the Casimir energy (expansion energy) at the Hubble scale (cosmic apparent horizon): 
the usual one $\sim A/4$ accounts for the remaining energy components of the universe.\\
However, 
a more precise calculation must be done by considering that the formula (\ref{termo}) has been derivated in a flat universe, while the
arguments of \cite{bb3}  do apply to a closed universe. For a closed universe \cite{14,15,17}, instead of (\ref{3}) the following 
holds:
\begin{equation}
\delta M\frac{G}{c^2}<\frac{L}{2}+\frac{AH}{4\pi c}-\frac{3V}{8\pi a(t)^2}.
\label{closed}
\end{equation}
Note that in (\ref{closed}) the spatial dimensions are due to $a(t)$.
It is simple to verify that, under the condition that $\{L,V/(a(t)^2)\}>c/H$
(strong self-gravitating regime), 
the first and the third therm on the right side of (\ref{closed}) are subdominant with respect to the 'Casimir like' one. 
Hence, by considering the suitable
normalized dynamical Bekenstein bound in this case, we obtain 
$S_{BH}=k_B HV/(2cL_P^2)$. It seems 
difficult to believe that this is only a coincidence.
Also note that the first and third term in (\ref{closed}) 
are vanishing by taking $V=4/3\pi R^3a(t)^3, L=Ra(t)$.   
Moreover, in the opposite regime (weak self-gravitating regime), or by considering 
black holes smaller than the 
Hubble radius, the leading term in
(\ref{closed}) is equivalent to the condition (\ref{3}), with $L=a(t)R(t)+o(1), V=4/3\pi R^3a(t)^3+o(1)$,
i.e. on sufficiently small scales, the negative curvature can be neglected.  

We do not go further in this analogy, but this indicates that we are on the right road.
 
\subsection{Equation of state for black holes in a flat expanding universe}

In this section, we analyze some consequences of (\ref{termo}) in relation to the equation of state of the black hole.
By differentiating (\ref{termo}) we get
\begin{equation}  
dS_{BH}=\frac{k_B}{4L_P^2} dA+\frac{3k_B}{2c L_P^2} V dH+
\frac{3k_B}{2c L_P^2}H dV.
\label{termo2}
\end{equation}
The first two terms in the right side of (\ref{termo2}) can be interpreted as representing $1/T$ times
the internal energy of the black hole.
In particular, the one  involving $dH$ can be seen as a variation 
of the internal energy 
due to the expansion of the universe caused by the presence of some unspecified kind of matter.
In a physically viable usual cosmology (i.e. satisfying weak and strong energy conditions)
we have $dH\leq 0$. Hence, in an expanding universe this term lowers the internal energy.
The term proportional to $dV$ can be seen as a work
term $dW$ due to the Hubble flow: for an expanding universe ($H>0$),
when $dV>0$  we have $dW>0$, while when $dV<0$ obviously $dW<0$, a reasonable fact. The introduction of a term proportional to 
$V$ in (\ref{termo2}) confirms that holographic principle struggles with cosmology and that more degrees of freedom are necessary, in particular the ones related to the non static nature of the spacetime.\\
Note that  by using theorems present in \cite{13,17},
the expression (\ref{entr}) can also be generalized to a Friedmann universe 
with negative curvature: in this case another positive term arises proportional to $\sim H^2A^2$ and in addiction more involved 
expressions for $L$ and $A$ arise.

Another interesting consequence of the modified Bekenstein-Hawking formula is the possibility to write down, thanks to the volume 
term $dV$ in (\ref{termo2}), an equation of state for the embedded black hole.
In fact, we can write:
\begin{equation}  
\frac{P}{T} =\frac{3k_B H}{2c L_P^2},
\label{w1}
\end{equation}
where $P$ denotes the pressure. For a universe with negative curvature, a further positive term proportional to $V/a(t)^2$ also appears in 
(\ref{w1}). The ratio $P/T$ measures the entropy variation with respect to a variation of the volume $V$. For a vanishing 
$H$, it follows that $P=0$.
Hence, a non-vanishing
pressure for a black hole can be an indication of the non static nature of the universe.\\ 
By denoting with $R_H$ the Hubble radius,
we can write formula (\ref{w1}) in the following form:
\begin{equation}  
P R_H L_P^2=\frac{3}{2}k_B T.
\label{ig}
\end{equation}
For an ideal gas we have $PV=N k_B T$, where $N$ is the particles number. 
It is interesting that in the formula (\ref{ig}) explicitally emerges the apparent horizon. 
After multiplying both members of
(\ref{ig}) for the proper volume $V$ of the apparent horizon of the black hole, we have:
\begin{equation}  
P V=\left(\frac{3V}{2 R_H L_P^2}\right)
k_B T.   
\label{ig2}
\end{equation}
Suppose now to decompose the effective
proper volume $V(t)$ of the apparent horizon in $n(t)$ elemenatary spherical cells of fixed proper radius $L_P$, i.e.
$V=4/3\pi n(t)L_P^3$. This can be justified, has remarked in section 5, 
in light of the papers \cite{v1,v2} in the context of a quantum non-commutatice spacetime
at the Planck length, where a minimal volume arises as a consequence of well motivated space-time uncertainty relations.
Then expression (\ref{ig2}) becomes:
\begin{equation}  
PV=\left(2\pi\frac{L_P}{ R_H}\right) n(t) k_B T.
\label{cb}
\end{equation} 
The expression (\ref{cb}) generalizes the usual equation of state
suitable for ideal gases in the context of black hole thermodynamics in Friedmann flat expanding spacetimes.  Formula 
(\ref{cb}) can have a fine physical interpretation. When the universe is cold, as at present time, many quantum degrees of freedom 
are frozen: this is described by the actual very low ratio of $L_P/R_H$.
But when the universe has been hot, more and more degrees of freedom have been 
excited ($L_P/R_H\sim 1$). 
At the Planck epoch, when $R_H\sim L_P$, we have $PV\sim n k_B T$, with $2\pi$ a geometric factor due to the 
sphericity of the black hole (there is not reason to put a sphere into a rectangular box). A similar phenomenon happens for the 
ordinary statistical mechanics. Note that $n(t)$ is an integer and so it is $R_H/L_P=N_P(t)$, provided that the length $R_H$ is
expressed in terms of the Planck length $L_P$.
Hence the equation (\ref{cb}) can be written in the
expressive form:
\begin{equation}  
PV=\left(2\pi\frac{n}{ N_P}\right)k_B T,
\label{trek}
\end{equation} 
a kind of Bohr-Sommerfield quantization rule for the black hole equation of state. We do not speculate further on this formula.

The presence of a volume term in (\ref{termo2}) could struggle with the holographic principle. However, note that
the work term $PdV$ of usual thermodynamics arises thanks to a volume dependence of the entropy, and is unavoidable.
Otherwise, no work would be associated to the expansion of $A$, that seems in a cosmological context rather unlikely.

It should also be noticed that the theorem (\ref{3}) remains valid if we substitute the energy-density
${\overline{\rho}}_m$ with a constant energy-density, i.e. in a de Sitter expanding universe. In such a case, we have
again the formula (\ref{termo}), but with $H(t)$ the constant de Sitter value
$H=\sqrt{\Lambda/3}$. In this case, the term involving $dH$ in (\ref{termo2}) is vanishing, and as a result the internal energy of a 
black hole in a de Sitter universe is left unchanged with respect to the asymptotically flat case. Equation 
(\ref{w1}) it gives $P/T\sim \sqrt{\Lambda/3}$. 

\section{Entropy Bound and Its Possible Generalization}

First of all, note that our formula (\ref{termo}) reduces, in a static background ($H=0$),
to the usual black hole one. In this context, the 
entropy bound $S_A(V)\leq k_B A/(4L_P^2)$ can be violated, where $S_A(V)$ denotes the entropy within a region of volume $V$
enclosed by $A$. The argument is the following. Suppose (see for example \cite{bb3}) to have a spherical star with entropy $S_u$ collapsing to form a black hole. For the generalized second law, the black hole will have an entropy $S_{BH}\geq S_u$. 
By following the star inside 
its own horizon, thanks to the spacelike singularity, 
its surface area $A$ will approach zero 
in a finite time but its entropy is at least $S_u$ and as a result the Bekenstein bound is violated.
This simple example, with the help of theorem (\ref{3tris}), can give some hint to a better formulation of the entropy spherical bound.
Consider the entropy bound in this formulation:
 
{\it In a given background, the entropy of a matter system enclosed in a compact hypersurface of
volume $V$ and area $A$ cannot exceed the entropy of the biggest  black hole formed if
a proper mass excess where present within $V$ 
violating the (\ref{3})}. 

Note that with respect to this formulation, the area $A$ of this new entropy bound does not refer necessarily
to the area of the hypersurface enclosing 
the system. This is an important difference that is in agreement with the phylosophy of the covariant Bousso bound. In
the example above, the biggest black hole is given by the one 
formed by the collapse of the star itself and the bound it is not given by $1/4$ the area of the collapsing object also within the horizon,
but rather from $1/4$ the area of the horizon in Planck units that thanks to the generalized second law cannot be violated. This also 
applies to a collapsing spherical shell within an existing black hole of area $A_h$, 
since there the 'biggest black hole' is the one just existing with horizon,  thanks to the contribution of the infalling shell,
greater than $A_h$.

Other examples violating the bound can be built by considering a system with a huge entropy. Also in this case, the judicious use of 
relation (\ref{3}) can help us to solve these situations.\\ 
Consider, in an asymptotically flat spacetime, a system made of a spherical ball of 
gas made of $N$ non interacting elements of rest mass $m_n$. Its entropy can be approximated  (also in a cosmological context see
\cite{bb5}) by $S_N=k_B N$. The (spacelike) entropy bound is violated iff:
\begin{equation} 
N>\frac{A}{4L_P^2}.
\label{vb1}
\end{equation}
Suppose now that the ball is composed of photons with wavelength $\lambda$ and energy 
$E=\frac{\hbar c}{2\pi\lambda}$. From (\ref{3}) we see that the condition under which a black hole do not form is 
$N<\pi L \lambda/L_P^2$. By setting the obvious inequality $\lambda\leq L$ we obtain 
$N<A/(4L_P^2)$. Therefore, also for the huge entropic photons ball, the (spacelike) entropy bound cannot be violated.

The situation is much more involved in a cosmological context. In fact, since the universe is filled with a non vanishing matter-energy
tensor $T_{\mu\nu}$, we can calculate the entropy content of a given closed region of proper area $A$ and volume $V$. In this context,
the dinamical equilibrium is obtained between the background density $\rho(t)$ and the Hubble flow $H(t)$: the Friedmann 
equations dictate  that an higher density 
$\rho$ implies an higher Hubble flow ($H^2\sim\rho$). Hence black holes, as well know, do not exist in Friedmann universes, also 
near the big bang where $\rho\rightarrow\infty$: to build a black hole, a perturbation with a positive mass excess violating
the (\ref{3}) must be imposed.\\
Consider a flat Friedmann universe together with a
spherical  homogeneous hypersurface 
at fixed time $t$.\\ 
For the volume and area we have $V=4/3\pi L^3,\;\;A=4\pi L^2$. The entropy $S_m$ of this region
can be written as \cite{bb2} $S_m=k_B\sigma V$, where $\sigma=S/a(t)^3$ and $S$ is the space and time constant comoving entropy depending generally on the mass-energy content within the comoving region enclosed by $V$.
The spacelike bound with the old proposal $S_ {BH}\sim A/4$ is violated iff:
\begin{equation}
L>\frac{3}{4\sigma L_P^2}.
\label{vb2}
\end{equation}
The inequality 
(\ref{vb2}) is satisfied for very large volumes 
and sufficiently near the big bang \cite{bb6} for causal usual fluids ($\alpha\geq 1/3$) below the particle horizon.
In our frame, the (\ref{vb2})  becomes:
\begin{equation} 
\sigma V_s>\frac{A_h}{4L_P^2}+\frac{3}{2L_P^2}V_h H.
\label{vb3}
\end{equation}
Note that in (\ref{vb3}) do appear the proper dimensions of the black hole with mass excess violating the (\ref{3}) and not the one of
{\bf s}. On general grounds we expect that $L_h\geq 2G M/c^2$ and thus 
\begin{equation} 
L_h\geq L_s+2L_s^2\frac{H}{c}.
\label{vb4}
\end{equation} 
Also by taking the equality relation for $L_h$ in (\ref{vb4}), in the (\ref{vb3}) do appear  terms containing the volume $V_s$ that can 
'struggle' with the volume term $\sigma V_s$. To this purpose, consider the more restrictive case by putting
$A_h=0$ in (\ref{vb3}) and $L_h=L_s$ in (\ref{vb4}),  after denoting with $t_P$ the Planck time,
the  (\ref{vb3}) becomes:
\begin{equation}
S>\frac{3\alpha}{2 c L_P^2 t_P^{\alpha}}t^{3\alpha-1},
\label{vb5}
\end{equation}
where we have fixed the scalar factor to be $a(t_P)=1$.
Note that, thanks to our added term only in the entropy formula, 
the comoving length of the hypersurface (two sphere) {\bf s} does not appear explicitely 
but only the constant entropy density $S$. However, 
this  dependence can be contained in $S$ and thus it 
must be verified if this dependence agrees with Einstein's equations predictions
or if
(\ref{vb5}) it is satisfied for a region larger than the dimensions of the 
observable universe.  To be more quantitative, consider once again
$S_m=k_B N_s(t)=k_B \sigma(t)V_s(t), \sigma(t)=N_s/V_s=\rho/m(t) $, where $m$ denotes the rest mass of the particles filling the universe that can also be a time function. This case is more general than the adiabatic case that can be regained by setting
$N_s(t)=const$. 
Hence, by using
the Einstein's equation $H^2=8/3\pi G{\rho}$, the inequality $S_m>k_B \frac{3}{2L_P^2}V_s H$ (restrictive hypothesis)
with $E=mc^2$ becomes:
\begin{equation}  
H>\frac{4\pi G E}{c^3L_P^2}.
\label{vb6}
\end{equation}
The most favorable situation can obtained with photons of proper wavelength ${\lambda}_p$. 
We finally get: $H>2c/{\lambda}_p$.
By taking  ${\lambda}_p=c/H$ (Hubble radius of the universe), we obtain $1>2$,
a contradiction. The adiabatic case is obtained as a subcase with $N_s(t)=const$ and
$E\sim 1/a(t)$ for photons in an expanding universe.
This simple example shows as the added term works.
The use of the Hubble radius is justified from the fact that it is an apparent cosmological horizon.
It is interesting to note that, with the use of the old expression $S_{BH}\sim A/4$, we obtain that the bound is exactly saturated 
at the Hubble radius $c/H$, but it is violated for larger regions within the particle horizon. Conversely, with our added term only, 
the bound is exactly saturated at the particle horizon, a reassuring fact. This certainly implies that our suggested improvement of the
entropy bound certainly works and that perhaps also a stronger version ($L_s\sim L_h$) can be suitable.

Note that we have considered only the spherical case. 
However, our proposal for a generalized bound can also be formulated for 
non-spherical black holes. In fact, we can advantage of the isoperimetric Penrose inequality 
(see \cite{v1,v2} and references therein) and its generalization \cite{v2} in a cosmological context to justify our expression
(\ref{termo}) also for non spherical configurations.

As a final consideration of this section, note that all the reasonings above are valid until
a classical description of the geometry is available. At the Planck scale, a classical description of the geometry is certainly
uncorrect. Suppose to explore the quantum world.
To this purpose, consider the original spacelike entropy bound applied to photons within an enclosed volume.
By using the Heisemberg uncertainty relation, 
we obtain $A\geq 4\pi L_P^2$ and hence a minimal surface (and also volume) does appear. 
This can be obtained in the context of a non-commutative geometry \cite{v1,v2},  also for non-spherical volumes,
leading to a quantum spacetime at the Planck length. The existence of a minimal surface and a minimal volume could be the true 
reason for which a limit exists on the informations that a certain region can contain. Moreover, in an expanding universe, due to the Hubble flow, this minimal surface (and volume) becomes smaller by using theorem (\ref{3}) (see \cite{v2}), allowing 
more information (entropy) to be stored within a given surface, according to our formula (\ref{termo}).

\section{Primordial Inflation from Black Hole Evaporation?}
The idea that black hole physics could inspirate an alternative mechanism to begin inflation is not new 
(see for example \cite{infl1,infl2}). In particular in \cite{infl1} the apparent horizon is considered as a valid object capable  to produce an 
ingoing radiation similar to the Hawking one allowing inflation.\\ 
In \cite{infl2} micro black holes remnants can induce a matter dominated universe before the inflation.\\ 
In this subsection we shortly explore the possibility that primordial black hole evaporation 
can provide a mechanism to begin primordial inflation. As a first step, suppose that the primordial inflation soon after the Planck era is
dominated by a foam of micro black holes. 
As a title of example, we consider a single black hole with size $L$ that can be greater equal or less than 
the Hubble radius $c/H$. 
As shown in \cite{infl1}, this universe can be considered adiabatic and open since of the evaporation process. Contrary to the 
mechanism shown in \cite{infl1}, the Hawking radiation born from black hole evaporation produces an outgoing flux.
Since for an adiabatic universe $TdS_{BH}=dQ$,  $V=4\pi/3 L^3, A=4\pi L^2$ (i.e. $L$ is the proper radius of the apparent horizon)
and after deviding for $dt$ we obtain (see \cite{infl1}):
\begin{eqnarray}
& &\frac{\pi k_B}{L_P^2}2L L_{,t}+\frac{2\pi k_B}{cL_P^2}L^3H_{,t}+\frac{6\pi k_B}{cL_P^2}HL^2L_{,t}=
-P_{rad},\nonumber\\
& &P_{rad}=\frac{{\pi}^2k_B^4}{60c^2{\hbar}^3}A T^3,\;\;T=\frac{\hbar c}{2\pi k_B L}
|1-\frac{L_{,t}}{2H L} |. \label{i1}
\end{eqnarray}
The second and third term at the right hand side of (\ref{i1}) are a consequence of our modification of the original Bekenstein-Hawking
entropy.
In the following we investigate the viability that the equation (\ref{i1}) allows an inflationary quasi de Sitter phase. Soon after the Planck era the universe expanded very quickly. Therefore we may reasonably suppose that in such a phase $L_{,t}>0$, i.e. that the black holes are 
influenced by the tremendous Hubble flow. 
However, as the time
flow, at early times, 
the expansion rate decreases and the evaporation mechanism
becomes more and more efficient. As a consequence, there existed 
an early phase where
$L_{,t}\simeq 0, L_{,t}>0$ and $H_{,t}\simeq 0$. For $H_{,t}>0$ we have a blue-shifted expansion while for 
$H_{,t}<0$ we have a red-shifted inflation. The  Planck data \cite{plan} predicts a red-shifted inflation, while the more recent BICEP2
one \cite{Bic} could account also for 
a blue-shifted inflation \cite{mad}. However the combination of Planck and BICEP2 data is still in good agreement with a red-shifted inflation.\\
First of all, note that in this phase certainly $|L_{,t}/L|<<H$ and hence 
$T\sim 1/L$. 
Suppose that at the beginning of the inflationary phase $H_I\geq c/L_I$, 
where the pedix '$I$' denotes the value at the starting of inflation.
Also suppose that  $|H_{,t}/H|>> |L_{,t}/L|$ (which implies a very small expansion rate for the apparent horizon of the micro black hole)
and hence the second term in the left 
hand side of (\ref{i1}) dominates:
as a consequence from (\ref{i1}) we obtain:
\begin{equation}
H=H_I-\frac{c^2L_P^2}{240\pi L_I^4}\left(t-t_I\right)+o(1),
\label{i2}
\end{equation}
i.e. a red-shifted inflation. The ratio $L_P^2/L_I^4$ measures the non gaussianity level of the perturbations. 
Note that the presence of the factor $L_P^2$ in (\ref{i2})  can hint a possible non-commutative 
signature of the inflation. In the limit $L_P=0$ the evolution is purely de Sitter.
Since 
we expect that $\Delta t=t_F-t_I\in[10^{-34},10^{-32}] s$ ($t_F$ denotes the end of the inflation), we must have 
\begin{equation}
H_I>>\frac{c^2L_P^2\Delta t}{240\pi L_I^4}.
\label{hi}
\end{equation}
In the opposite regime, i.e. $|H_{,t}/H|\leq |L_{,t}/L|$ and $H_I<<c/L_I$, we obtain 
\begin{equation} 
L^3=L_I^3-\frac{cL_P^2}{720\pi}\left(t-t_I \right)+o(1).
\label{i3}
\end{equation} 
We obtain a monotonically decreasing expression for $L$ during inflationary epoch. This solution struggles with the 
reasonable hypothesis that $L_{,t}>0$ just before inflation. If we dismiss this assumption, solution 
(\ref{i3}) can lead to a blue-shifted inflation by considering, for example $H\sim c/L$.

These preliminary crude estimations indicate that black hole inflation as a viable possibility. Apart from equation
(\ref{i1}), we must also consider the Einstein's equations. In particular, we can consider a universe filled with a foam of primordial black holes
that can be modelized as dust with density ${\rho}_{mbh}$ (see \cite{infl2}) together with the radiation ${\rho}_{rad}$. The Einstein's equations are thus $H^2=8/3\pi G\left({\rho}_{mbh}+{\rho}_{rad}\right)$ together with the continuity equations for 
${\rho}_{mbh}$ and ${\rho}_{rad}$. 
This is certainly matter of further investigations.

\section{Conclusions}
In this paper we have considered the Bekenstein-Hawking entropy formula in flat expanding universes in relation to known theorems of the 
general relativity for the formation of trapped surfaces. In particular, we quoted the theorems present in 
\cite{14,15,17}. Thanks to these theorems, 
we are in the position to propose a generalization of  the Bekenstein-Hawking entropy in a flat expanding
universe. The expression contains an extra term with respect to the usual expression taking into account the non static
nature of a cosmological background represented by the hubble flow $H$.\\ 
This new proposal for $S_{BH}$
allows us to write the equation of state of a black hole (equation (\ref{cb})). Interestingly enough, this equation of state depends on the ratio
$L_P/R_H$, that measures the deviation from the usual equation of state for an ideal fluid. This factor becomes of the order of unity 
when $L_P\sim R_H$, i.e. near the Planck era.

Our approach can be also extended to Friedmann universes with negative and positive curvature with embedded a black hole
(although in the case of closed universes it is not yet clear how to define a black hole). The only difference with the flat case
concerns the expressions for the effective volume, proper length and proper area of the black hole, where geometrical factors do appear.

In this paper we have considered spherically symmetric configurations. Hence, the proper area $A$ and the proper effective volume
$V$ are not independent quantities. Nevertheless, the identification  $A^{3/2}\sim V$ in (\ref{entr}) is certainly well motivated. 
If where 
possible to consider the expression (\ref{termo}) for non spherical configurations
(Kerr black holes embedded in Friedmann universes), then we could consider transformations with
$dA=0$ and $dV\neq 0$ that
leave the old entropy formula $S_{BH}\sim A/4$ unchanged. Conversely, in our context we have an effective term $\sim VH$ that is not
left unchanged by an isoperimetric transformation. Fortunately,  the theorem leading to inequality (\ref{3bis})
has been generalized in \cite{15} to non-spherical configurations with the condition that $S$ be an equipotential convex surface,
allowing to write for $S_{BH}$ an expression closed to the  formula (\ref{termo}), but with the difficulty
to introduce a good definition for the  volume of a non spherical black hole. In any case, we point out this important conceptual difference
with the old proposal.

We also explore the possibility that primordial inflation can be enhanced by black hole evaporation. Simple preliminar computations show
that this possibility is compatible with our equations. In particular, it is thanks to our added term that red-shifted inflation can be easily
obtained.
This can be certainly matter for further investigations.

As a final remark, note that an higher value for the entropy with respect to the usual formula in expanding universes implies that it is possible
to store more informations (remember that $I=S/\ln(2)$) in a given region than the ones contained in its surface and thus that holographic principle is not enough for a suitable knowledge of our universe.

\section*{Acknowledgements} 
I would like to thank Alessandra D'Angelo for interesting discussions and suggestions,
Luca Tomassini for many useful discussions on the quantum nature of the spacetime.
Finally, I also thank  the interesting comments of the 
anonymous referee that contributed to an improvement of this paper.

\end{document}